# Healthy Access for Healthy Places: A Multidimensional Food Access Measure

Irena Gao[1], Marynia Kolak[2,*]

[1] *Illinois Mathematics and Science Academy, Aurora, IL*
[2] *University of Chicago, Chicago, IL*

**Abstract**
When it comes to preventive healthcare, place matters. It is increasingly clear that social factors, particularly reliable access to healthy food, are as determinant to health and health equity as medical care. However, food access studies often only present one-dimensional measurements of access. We hypothesize that food access is a multidimensional concept and evaluated Penchansky and Thomas's 1981 definition of access. In our approach, we identify ten variables contributing to food access in the City of Chicago and use principal component analysis to identify vulnerable populations with low access. Our results indicate that within the urban environment of the case study site, affordability is the most important factor in low food accessibility, followed by urban youth, reduced mobility, and higher immigrant population.

## 1. Introduction

*1.1 Purpose*

When it comes to preventive healthcare, place matters. Health is determined as much by the individual's social and economic environment as by medical care. Without healthy nutrition, diabetes cases increase; without safe housing, lead poisoning increases. Public health policy has recently realized that it must ensure healthy social ecologies to nurture healthy people (Barnes 2009).

Access to healthy resources, especially food, is strongly correlated with safe social determinants of health. Preliminary studies on disparities in health access suggest that many of the dramatic health problems we face can be mitigated by careful policy that improves access. However, current access studies often incorporate simplified methodologies that only consider limited dimensions of access. The quantity of inconclusive and insignificant results saturating the field suggests that food access lacks an underlying theory (Caspi, Sorensen, Subramanian, & Kawachi, 2012; Charreire et al., 2010).

We evaluate a proposed multidimensional food access measure as modeled after Penchansky and Thomas (1981) to answer two critical questions: (1) *do independent dimensions of food access exist?* and, if so, (2) *which dimensions affect food access most?*

*1.2 Goals and Objectives*

The goals of this study are to:
(1) Determine whether independent dimensions of food access, as modeled after Penchansky and Thomas (1981) independently exist.
(2) Determine the relative strength of the different food access dimensions.

## 2. Literature Review

Medical care alone is not enough for health. The rise of chronic diseases suggests that many of today's health crises are tied to lifestyle, opportunity, and environment. Of the top ten causes of 2014 U.S. deaths in the Center for Disease Control's (CDC) *Health, United States 2015* report, six (*i.e.* heart disease and diabetes) are more closely tied to how we live than even genetics (Ludwig et al., 2011; National Center for Health Statistics, 2016). Healthy people must live healthy lifestyles—and healthy lifestyles are tied to the healthiness of the individual's social ecology. Just as septic housing invited cholera during the Industrial Revolution, healthy neighborhoods affect healthy food consumption and activity, leading to lower obesity rates (Booth et al., 2001; Ludwig et al., 2011). As part of its Healthy People 2020 objective, the Office for Disease Prevention and Health Promotion has recognized the policy responsibility to improve these social determinants of health (Orszag, Barnes, Carrion, & Summers, 2009), which may also be the key to improving health equity, as racial and socioeconomic health disparities are majorly impacted by environmental inequalities (Brulle & Pellow, 2006).

When we consider social factors of health, a repeated theme is the idea of equal and adequate access. In particular, healthy food access is critical for health. Various studies have suggested that poor access to healthy food providers correlates with poor diet and, consequently obesity (Walker, Keane, & Burke, 2010). Methodologies in the field have little standardization, spanning quantitative Geographical Information Systems (GIS) and qualitative food store assessments, surveys, and focus groups (Caspi, Sorensen, et al., 2012; McKinnon, Reedy, Morrissette, Lytle, & Yaroch, 2009; Walker et al., 2010). In general, GIS-methods dominate the field, outnumbering qualitative studies 57 to 10; because of their affordability and speed, these methods will only increase in popularity (Charreire et al., 2010).

Recently, food access methodologies, particularly GIS systems, have been critiqued for returning different and sometimes incompatible results (Caspi, Sorensen, et al., 2012; Charreire et al., 2010). One study found discordance between perceived access, GIS measures of access, objective walking access, and actual fruit and vegetable intake. Notably, Boston area residents who lived in accessible regions by GIS measures were the ones who reported food insecurity and had low produce intake (Caspi, Kawachi, Subramanian, Adamkiewicz, & Sorensen, 2012). Even within GIS methodologies, different measures of access (*e.g.* container, gravity potential, travel cost, and minimum distance) return distinct, if not contradictory, results because the GIS measure used makes assumptions on the situation being studied; accurate GIS measures must reflect the actual underlying dynamics (Ball, Timperio, & Crawford, 2009; Talen & Anselin, 1998). Several phenomena suggest that research design, especially in GIS studies, is critical to achieving usable results. Individual GIS-based studies are sometimes inconclusive; in one study, proximity to fast food stores decreased the chance of Australian children eating fruit more than 2 times a day, but the likelihood of children eating fruit more than three times a day increased as distance to supermarkets *increased* (Timperio et al., 2008). Even when GIS-based results are conclusive, they tend to be insignificant. For example, one study found that a one mile increase in proximity to a supermarket was statistically significant but only associated with a 0.02 difference in produce servings per day (Sharkey, Johnson, & Dean, 2010). These studies suggest that GIS tools may be often misused when not informed by accurate research design informed by sound theory.

The pressing question in the food access field, then, is "why?". Why do different methodologies for access so ambiguously vary? A small body of research has suggested that the fundamental problem of the field is the negligence of a formal model of access (Caspi, Sorensen, et al., 2012; Charreire et al., 2010). Indeed, different methodologies appear to have unambiguous differences in the *aspects* of access they address. Several studies have found that non-spatial factors, especially affordability, may affect access more than spatial measures of availability (Donkin, Dowler, Stevenson, & Turner, 1999; Drewnowski, Aggarwal, Hurvitz, Monsivais, & Moudon, 2012). In 2012, Caspi suggested the formation of a multi-dimensional model of food access based on Penchansky and Thomas's five-prong model of healthcare access, which includes availability, accessibility, affordability, acceptability, and accommodation (Penchansky & Thomas, 1981). While this model has yet to be critically examined in the field, the adaption of this model offers a valuable opportunity to examine two critical questions: (1) *do independent dimensions of food access exist?*, (2) *are they spatial*?, and (3) *which dimensions affect food access most?*

## 3. Methodology

*3.1 Overall Approach*

In our study, we use spatial GIS tools. In consideration of the methodology concerns reviewed earlier, we enforce our methodology with a theoretical model to reflect a hypothesized underlying model of the food environment's dynamics. Based on suggestions from Caspi (2012), we examine the Penchansky and Thomas model for food access via spatial data analysis. Of these five dimensions, the first two are most familiar in GIS methods. *Availability* refers to the presence of food supplies; *accessibility* refers to the ease of locating and getting food supply. Only a small body of literature discusses *affordability* of food prices. *Acceptability* refers to people's attitudes about their food environment. *Accommodation* refers to how well food suppliers adapt to residents' needs, such as acceptance of food stamp assistance or non-English speakers.

We examine ten variables associated with four of the Penchansky and Thomas dimensions (Table 1). Variables were at the census tract resolution for the 791 census tracts in Chicago, Illinois. We did not examine concrete variables for acceptability, as acceptability is, by definition, a strictly subjective dimension incompatible with quantitative analysis. Instead, any inability of our variables to explain access variation may suggest influence by the acceptability dimension.

The AV_INT variable (Table 1) is calculated using QGIS's variable buffer and intersection analysis. The Chicago Data Portal grocery store data is notable for its inclusion of non-chain grocery stores. Using exploratory spatial data analysis in GeoDa, we first examine spatial autocorrelation of each variable and correlation between variables. Next, we use multivariate analysis in R to evaluate the relative strength of each variable to make suggestions on the validity of a Penchansky and Thomas food access model.

*3.2 Data*

| Table 1. Variables and Data Sources | | | |
|---|---|---|---|
| Variable | Description | Penchansky & Thomas Proposed Dimension | Data Source |
| AV_INT | Count of intersections with buffer zones of food providers<br>• Supermarkets (3km)<br>• Grocery stores (1.6km or 0.8km)<br>• Produce carts (0.5km)<br>• Farmers markets (1km) | Availability | Supermarkets Grocery stores: Chicago Data Portal Farmers markets: Chicago Data Portal Produce carts: Chicago Data Portal |
| AV_POP | Population density of tract | Availability | 2015 American Community Survey 5-year average, as pulled from the 2014 Social Vulnerability Index |
| ACE_NET | Average network distance to nearest supermarket via residential roads | Accessibility | mentor-led research project (Kolak: Northwestern, |

| | | | University of Chicago, Chicago State University) |
|---|---|---|---|
| ACE_NV | % of households without a vehicle | Accessibility | 2015 American Community Survey 5-year average, as pulled from the 2014 Social Vulnerability Index |
| ACE_ELD | % of population age 65+ | Accessibility | 2015 American Community Survey 5-year average, as pulled from the 2014 Social Vulnerability Index |
| ACE_DIS | % of population disabled | Accessibility | 2015 American Community Survey 5-year average, as pulled from the 2014 Social Vulnerability Index |
| AFF_POV | % of population below the poverty line | Affordability | 2015 American Community Survey 5-year average, as pulled from the 2014 Social Vulnerability Index |
| AFF_UNEMP | % of population unemployed | Affordability | 2015 American Community Survey 5-year average, as pulled from the 2014 Social Vulnerability Index |
| ACO_ENG | % of population speaking English "less than very well" | Accommodation | 2015 American Community Survey 5-year average, as pulled from the 2014 Social Vulnerability Index |
| ACO_SNAP | % population reporting regular use of SNAP, Chicago's food stamp program | Accommodation | 2015 American Community Survey 5-year average, as pulled from the 2014 Social Vulnerability Index |

## 4. Results

The results of the Principal Component Analysis are shown in Table 2. As the first four principal components explain 81.9% of the variance, we focus our discussion on the populations represented by these four components. We defined each component's significant contributing variables as variables whose loadings were at least 0.4000 in either direction and secondary contributing variables as variables whose loadings were at least 0.1000 in either direction.

| Table 2. PCA Results | | | | | | | | | | |
|---|---|---|---|---|---|---|---|---|---|---|
| | PC 1 | PC 2 | PC 3 | PC 4 | PC 5 | PC 6 | PC 7 | PC 8 | PC 9 | PC 10 |
| Proportion of Variance | 0.4068 | 0.2056 | 0.1127 | 0.0939 | 0.0498 | 0.0486 | 0.0294 | 0.0273 | 0.0141 | 0.0118 |
| Variable Loadings | | | | | | | | | | |
| AV_INT | 0.2829 | 0.3406 | 0.2905 | -0.0098 | 0.0999 | -0.8138 | -0.2051 | 0.0674 | -0.0216 | -0.0119 |
| AV_POP | -0.1845 | 0.5316 | 0.1506 | 0.0789 | -0.5102 | 0.2748 | -0.4744 | 0.1729 | 0.2120 | 0.1367 |
| ACE_NET | -0.2541 | -0.2855 | -0.4735 | -0.1457 | -0.6200 | -0.4622 | 0.0505 | 0.0437 | 0.0771 | -0.0113 |
| ACE_NV | -0.3167 | 0.1955 | 0.4965 | -0.2898 | -0.1618 | -0.0475 | 0.6843 | -0.0250 | 0.1720 | 0.0639 |
| ACE_ELD | -0.0998 | -0.4463 | 0.4630 | 0.4842 | -0.2021 | -0.0411 | -0.0085 | 0.4350 | -0.3256 | 0.0648 |
| ACE_DIS | -0.3744 | -0.2304 | 0.2442 | 0.3293 | 0.0861 | -0.1291 | -0.2388 | -0.5867 | 0.4109 | -0.2111 |
| AFF_POV | -0.4246 | 0.2851 | -0.0664 | -0.0194 | 0.0519 | -0.0131 | -0.0299 | 0.0102 | -0.5328 | -0.6678 |
| AFF_UNEMP | -0.4215 | -0.0074 | -0.1687 | -0.0066 | 0.4725 | -0.0989 | -0.0528 | 0.6128 | 0.4271 | -0.0086 |
| ACO_ENG | 0.1278 | 0.3538 | -0.3109 | 0.7315 | -0.0584 | -0.0503 | 0.4469 | -0.0051 | 0.1354 | -0.0522 |
| ACO_SNAP | -0.4435 | 0.1512 | -0.1367 | 0.1001 | 0.2039 | -0.1246 | -0.0400 | -0.2324 | -0.3995 | 0.6924 |

PC1 (40.68% of variance) represents a tract's economic prosperity. Vulnerable tracts have a low PC1 score (shown in blue in Figure 1). Affordability appears to be the main concern for these residents. Tracts have high poverty rates, high unemployment, and high SNAP usage. Availability of food providers and geographical accessibility are also concerns. These tracts are also more likely to have higher population density and have to travel further to fewer sores while not owning a vehicle and being disabled. Vulnerable tracts include the West and South Side of the city.

PC2 (25.06% of variance) represents a tract's composition of urban youth. Vulnerable tracts have a high PC2 score (shown in orange in Figure 2). Affordability is the major concern for these residents. These tracts are more densely populated and have fewer elderly residents, but are also more likely to have impoverished or non-English speaking residents using SNAP. On the other hand, food accessibility or availability does not appear to be a problem, as orange tracts have more stores, a shorter distance to stores, and are more likely to own a vehicle while not disabled. Vulnerable tracts are scattered across the center of the city.

PC3 (11.27% of variance) represents a population's innate mobility. Vulnerable tracts have a high PC3 score (shown in orange in Figure 3). While these tracts have a shorter network distance to stores and more vehicle ownership, residents are older. They are also more likely disabled in a population-dense area while being employed, English-speaking, and non-SNAP users. For these tracts, a mix of moderate accessibility risk appears to be the main concern. Vulnerable tracts cluster along the shoreline of the city.

PC4 (9.34% of variance) represents a tract's composition of immigrant families. Vulnerable tracts have a high PC4 score (shown in orange in Figure 4). For these tracts, ethnic barriers and affordability are the main concerns. These tracts have a higher proportion of elderly residents that depend on younger generations, while also speaking English poorly. They are more likely to be impoverished and use SNAP. On the other hand, they are less likely to have to travel further to a food provider and more likely to own a vehicle. Vulnerable tracts include the Northwest and West parts of the city, corresponding with high clusters of Hispanic, Polish, and Middle Eastern neighborhoods, respectively.

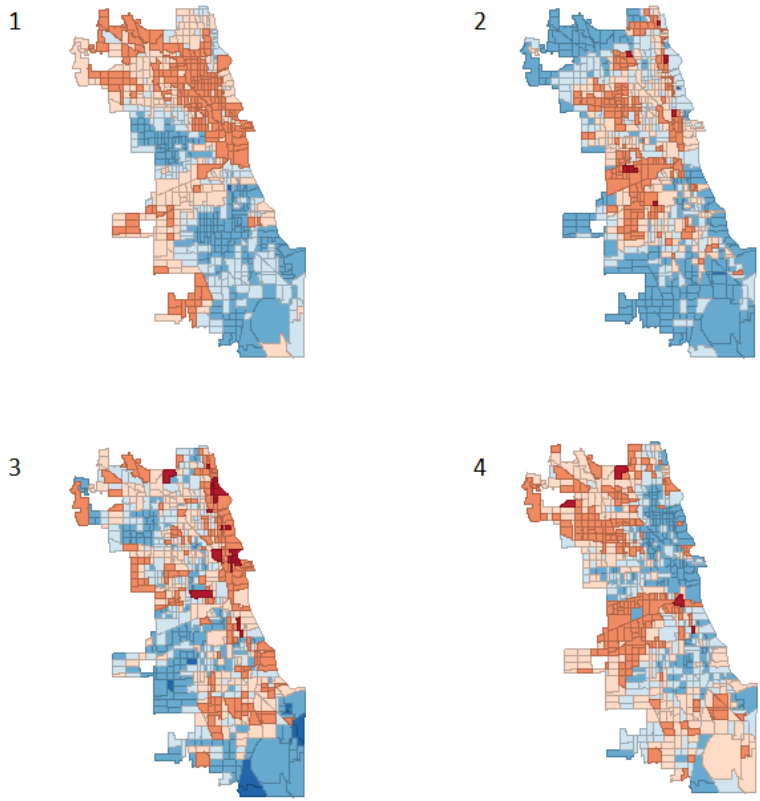

*Figures 1-4. Box maps (hinge = 1.5) of the PC1, PC2, PC3, and PC4.*

In addition, the correlation matrix between PC loadings is shown in Table 3. Of particular interest is the negative correlation between AFF_POV and ACE_NET, suggesting that if poverty increases in a vulnerable population, the network road distance to food providers decreases in importance, and the positive correlation between ACO_ENG and ACE_NET, suggesting that as English limitations increase, network road distance to food providers increases as a concern.

| Table 3. PCA loading correlation matrix | | | | | | | | | | |
|---|---|---|---|---|---|---|---|---|---|---|
|  | AV_INT | AV_POP | ACE_NET | ACE_NV | ACE_ELD | ACE_DIS | AFF_POV | AFF_UNEMP | ACO_ENG | ACO_SNAP |
| AV_INT | 1.0000 | -0.0007 | 0.0052 | -0.0015 | -0.0006 | 0.0014 | 0.0030 | -0.0015 | -0.0028 | 0.0004 |
| AV_POP | -0.0007 | 1.0000 | 0.1081 | -0.0311 | -0.0127 | 0.0281 | 0.0615 | -0.0301 | -0.0568 | 0.0090 |
| ACE_NET | 0.0052 | 0.1081 | 1.0000 | 0.2199 | 0.0899 | -0.1984 | -0.4341 | 0.2127 | 0.4009 | -0.0634 |
| ACE_NV | -0.0015 | -0.0311 | 0.2199 | 1.0000 | -0.0259 | 0.0571 | 0.1250 | -0.0612 | -0.1154 | 0.0183 |
| ACE_ELD | -0.0006 | -0.0127 | 0.0899 | -0.0259 | 1.0000 | 0.0233 | 0.0511 | -0.0250 | -0.0472 | 0.0075 |
| ACE_DIS | 0.0014 | 0.0281 | -0.1984 | 0.0571 | 0.0233 | 1.0000 | -0.1127 | 0.0552 | 0.1041 | -0.0165 |
| AFF_POV | 0.0030 | 0.0615 | -0.4341 | 0.1250 | 0.0511 | -0.1127 | 1.0000 | 0.1209 | 0.2278 | -0.0361 |
| AFF_UNEMP | -0.0015 | -0.0301 | 0.2127 | -0.0612 | -0.0250 | 0.0552 | 0.1209 | 1.0000 | -0.1116 | 0.0177 |
| ACO_ENG | -0.0028 | -0.0568 | 0.4009 | -0.1154 | -0.0472 | 0.1041 | 0.2278 | -0.1116 | 1.0000 | 0.0333 |
| ACO_SNAP | 0.0004 | 0.0090 | -0.0634 | 0.0183 | 0.0075 | -0.0165 | -0.0361 | 0.0177 | 0.0333 | 1.0000 |

## 5. Discussion

Our principal component analysis results do not strictly follow Penchansky and Thomas's model of five equal dimensions. Rather, our vulnerable populations vary in weight, or percent of variance explained. In addition, the first four vulnerable populations appear to have mixed Penchansky and Thomas concerns. We do confirm, however, that dimensions thought of as "non-spatial" (*i.e.* affordability) do actually follow spatial patters. Our PC1 "economic prosperity" variable best demonstrates this feature.

The consistent significance of age structure in our components, often separate from other accessibility variables, suggests that the appropriate dimensions of food access are highly dependent on the city's personal demographic and social structure. Food access does appear to have dimensions; however, those dimensions are specific to the populations of each individual city. Similar analyses should be conducted by city departments to identify their needs.

The correlation of our economic prosperity and immigrant family components with racial composition confirms existing food access research that the Chicago food environment falls along racial lines (Abrajano, Andrade, B. Martin, & Memon, 2011). In our study, we trace this apparent segregation to economic and demographic variables.

It is important to note that our study is not validated by health outcomes or survey results. In addition, our study may not have captured all possible significant variables. In addition to cross-checking our results with health outcomes and survey data, future validation could consist of tracking these variables across time to confirm prevalence.

In Chicago, at least, one concern stands out: food affordability. Secondary concerns are in the availability and accessibility of food. Our results suggest that Chicago food policy should consider efforts to improve food access that include targeting affordability, in addition to accessibility and availability dimensions.

## 6. Conclusions

This work presents a multivariate analysis of eleven variables about Chicago's food access environment, tentatively categorized according to Penchansky and Thomas's 1981 paper on healthcare access. Our analysis suggests that the demographic, historical, and social composition of cities result in individual food access dimensions in different weights, but the Penchansky and Thomas dimensions of affordability, accessibility, and availability are useful guidelines for articulating city needs. Within Chicago, significant attention is required to address food affordability.

Perhaps this study's main message can be summarized as a need for dialogue. The food access field's silence on establishing a conceptual model for food access risks neglecting other dimensions of access, especially affordability, that this study suggests are critical. Food access research has the potential to alleviate deepening food stress in communities. But for food access to take a step forward, food research must take a step back and ensure it has a solid theoretical foundation.